\documentstyle[12pt]{article}

\begin{document}
{\sf \begin{center} \noindent
{\Large \bf Twist, writhe and energy from the helicity of magnetic perturbed vortex filaments and AB effect}\\[3mm]

by \\[0.3cm]

{\sl L.C. Garcia de Andrade}\\

\vspace{0.5cm} Departamento de F\'{\i}sica
Te\'orica -- IF -- Universidade do Estado do Rio de Janeiro-UERJ\\[-3mm]
Rua S\~ao Francisco Xavier, 524\\[-3mm]
Cep 20550-003, Maracan\~a, Rio de Janeiro, RJ, Brasil\\[-3mm]
Electronic mail address: garcia@dft.if.uerj.br\\[-3mm]
\vspace{2cm} {\bf Abstract}
\end{center}
\paragraph*{}
The twist and writhe numbers and magnetic energy of an orthogonally
perturbed vortex filaments are obtained from the computation of the
magnetic helicity of geodesic and abnormal magnetohydrodynamical
(MHD) vortex filament solutions. Twist is computed from a formula
recently derived by Berger and Prior [J. Phys. A 39 (2006) 8321] and
finally writhe is computed from the theorem that the helicity is
proportional to the sum of twist and writhe. The writhe number is
proportional to the total torsion and to integrals of the vector
potential. The magnetic energy is computed in terms of the integral
of the torsion squared which allows us to place a bound to energy in
in the case of helical filaments due to a theorem by Fukumoto [J.
Phys. Soc. Japan,(1987)] in fluid vortex filaments. It is also shown
that filament torsion coincides with the magnetic twist in the case
under consideration, where a small orthogonal magnetic field exist
along the thin filament. A new Aharonov-Bohm (AB) phase term is
obtained in the writhe number expression which is not present in the
Moffatt-Ricca (Proc Roy Soc London A,1992)
computation.\vspace{0.5cm} \noindent {\bf PACS numbers:}
\hfill\parbox[t]{13.5cm}{02.40}

\newpage
\section{Introduction}
 The topology and geometry of fluid vortex filaments \cite{1} have immensely contributed to our comprehension of
 the \cite{2} dynamics and kinematics of magnetic filamentary structures  with applications in solar and  plasma
 physics \cite{3,4}. In particular twisted filamentary structure have helped solar physicists to better
understand the mechanism which allow the highly twisted coronal
magnetic flux tube to emerge from the solar surface and produce
those beautiful solar flares and loops. Recently Berger and Prior
\cite{5} have presented a detailed discussion of twist and writhe of
open and closed curves presenting a formula for the twist of
magnetic filaments in terms of parallel electric currents. In the
present paper we apply Berger-Prior formula and the theorem of the
sum of writhe and twist numbers investigated previously by Moffat
and Ricca \cite{6} and Berger and Field \cite{7} to obtain an
expression for the writhe number for the magnetically perturbed
vortex filamentary twisted structure. To obtain the writhe number we
make use of the helicity local expression for the magnetic field
proportional to the magnetic field itself and decompose the magnetic
vector field along the Serret-Frenet frame in 3D dimensions. This
allows us to solve scalar MHD equations to obtain constraints on the
global helicity expression which will allow us to compute the writhe
number. The knowledge of tilt , twist and writhe of solar filaments
for example has recently helped solar physicists  \cite{8} to work
out data obtained from the vector magnetograms placed in solar
satellites. This is already an strong motivation to go on
investigating topological properties of these filamentary twisted
magnetic structures. Earlier the Yokkoh solar mission has shown that
the sigmoids which are nonplanar solar filaments are obtained due to
the action of electric currents along these filaments which was used
recently \cite{9} as motivation to investigate current-carrying
torsioned twisted magnetic curves Throughout the paper we use the
notation of a previously paper on vortex filaments in MHD.
Mathematical notation is used based on the book by C. Rogers and W.
Schief \cite{10} on the geometry of solitons. The paper is organized
as follows: In section 2 we decompose MHD equations on a Frenet
frame along the twisted thin filament and solve the scalar equations
obtained. In section 3 we compute the twist, writhe and energy of
the magnetic filament. In section 4 we present the conclusions.
 \section{MHD scalar equations for twisted filamentary structures}
 Let us now start by considering the MHD field equations
\begin{equation}
{\nabla}.\vec{B}=0 \label{1}
\end{equation}
\begin{equation}
{\nabla}{\times}{\vec{B}}= {\alpha}\vec{B} \label{2}
\end{equation}
where ${\alpha}$ is the magnetic twist and the magnetic field
$\vec{B}$ along the filament is defined by the expression
$\vec{B}=B_{s}\vec{t}+B_{n}\vec{t}$ and $B_{n}$ is the magnetic
field perturbation orthogonal to the filament all along its
extension, and $B_{s}$ is the component along the arc length s of
the filament. The vectors $\vec{t}$ and $\vec{n}$ along with
binormal vector $\vec{b}$ together form the Frenet frame which obeys
the Frenet-Serret equations
\begin{equation}
\vec{t}'=\kappa\vec{n} \label{3}
\end{equation}
\begin{equation}
\vec{n}'=-\kappa\vec{t}+ {\tau}\vec{b} \label{4}
\end{equation}
\begin{equation}
\vec{b}'=-{\tau}\vec{n} \label{5}
\end{equation}
the dash represents the ordinary derivation with respect to
coordinate s, and $\kappa(s,t)$ is the curvature of the curve where
$\kappa=R^{-1}$. Here ${\tau}$ represents the Frenet torsion. We
follow the assumption that the Frenet frame may depend on other
degrees of freedom such as that the gradient operator becomes
\begin{equation}
{\nabla}=\vec{t}\frac{\partial}{{\partial}s}+\vec{n}\frac{\partial}{{\partial}n}+\vec{b}\frac{\partial}{{\partial}b}
\label{6}
\end{equation}
 The other equations for the other legs of the Frenet frame are
\begin{equation}
\frac{\partial}{{\partial}n}\vec{t}={\theta}_{ns}\vec{n}+[{\Omega}_{b}+{\tau}]\vec{b}
\label{7}
\end{equation}
\begin{equation}
\frac{\partial}{{\partial}n}\vec{n}=-{\theta}_{ns}\vec{t}-
(div\vec{b})\vec{b} \label{8}
\end{equation}
\begin{equation}
\frac{\partial}{{\partial}n}\vec{b}=
-[{\Omega}_{b}+{\tau}]\vec{t}-(div{\vec{b}})\vec{n}\label{9}
\end{equation}
\begin{equation}
\frac{\partial}{{\partial}b}\vec{t}={\theta}_{bs}\vec{b}-[{\Omega}_{n}+{\tau}]\vec{n}
\label{10}
\end{equation}
\begin{equation}
\frac{\partial}{{\partial}b}\vec{n}=[{\Omega}_{n}+{\tau}]\vec{t}-
\kappa+(div\vec{n})\vec{b} \label{11}
\end{equation}
\begin{equation}
\frac{\partial}{{\partial}b}\vec{b}=
-{\theta}_{bs}\vec{t}-[\kappa+(div{\vec{n}})]\vec{n}\label{12}
\end{equation}
Substitution of these equations into the magnetic helicity equation
reads
\begin{equation}
{\nabla}{\times}\vec{B}=\vec{t}[B_{n}(div\vec{b})-B_{s}({\Omega}_{s}+{\tau})]+\vec{n}{\tau}B_{n}+\vec{b}[{\kappa}B_{s}-B_{n}(1+{\theta}_{ns})]\label{13}
\end{equation}
while its RHS is
\begin{equation}
{\alpha}\vec{B}={\alpha}[B_{n}\vec{n}+B_{s}\vec{t}] \label{14}
\end{equation}
Comparing both sides component by component one obtains the three
scalar equations
\begin{equation}
{\tau}B_{n}={\alpha}B_{n} \label{15}
\end{equation}
This result seemed to be noticed earlier by Parker \cite{11} which
used to call the parameter ${\alpha}$ torsion. Throughout this
derivation we consider that the parameter ${\alpha}$ is constant,
which is not very usual in other works on magnetic helicity. The
other scalar equations are
\begin{equation}
\frac{B_{n}}{B_{s}}=\frac{\kappa}{(1+{\theta}_{ns})} \label{16}
\end{equation}
which is also usual when one deals with magnetic flux tubes
\cite{12} and finally
\begin{equation}
{\alpha}{B_{s}}=B_{n}div{\vec{b}}-B_{s}({\Omega}_{s}+{\tau})
\label{17}
\end{equation}
The equation ${\nabla}.\vec{B}=0$ becomes
\begin{equation}
{\partial}_{s}{B}_{s}+[{\theta}_{bs}+div\vec{b}]{B}_{s}=0 \label{18}
\end{equation}
A simple solution of this equation can be obtained if one considers
that $div\vec{b}$ does not vary appreciably on a short distance ds
along the filament. The solution is
\begin{equation}
{B}_{s}=B_{0}[1-div{\vec{b}}\int{{\kappa}ds}] \label{19}
\end{equation}
where $B_{0}$ is an integration constant. Substitution into equation
(\ref{17}) assuming that the flow is geodesic along the filament
and abnormality relation ${\Omega}_{s}=0$ along with
${\theta}_{bs}={\theta}_{ns}=0$one obtains
\begin{equation}
{B}_{n}=B_{0}[(div{\vec{b}})^{-1}-\int{{\kappa}ds}] \label{20}
\end{equation}
which allows us to say that we obtain a solution for the magnetic
filament in terms of the vortex filament invariant
$\int{{\kappa}ds}$. In the next section we shall compute the
magnetic energy in terms of the magnetic vector potential $\vec{A}$,
which allows us to compute the helicity integral and the writhe
number in terms of the AB phase.
\section{Magnetic energy, Twist and Writhe}
To be able to compute the magnetic energy , twist and writhe we now
compute the vector potential assuming that it obeys the Coulomb
gauge ${\nabla}.\vec{A}=0$ and the definition
$\vec{B}={\nabla}{\times}\vec{A}$. The Coulomb gauge becomes
\begin{equation}
{\partial}_{s}A_{s}+div{\vec{n}}A_{n}=0 \label{21}
\end{equation}
together with the equations for the definition of $\vec{B}$, taking
$\vec{A}=A_{s}\vec{t}+A_{n}\vec{n}+A_{b}\vec{b}$ yields
\begin{equation}
\frac{A_{s}}{A_{b}}=-\frac{\tau}{\kappa} \label{22}
\end{equation}
which was obtained from the constraint $B_{s}=0$ and
\begin{equation}
{B_{n}}=-{\tau}A_{n} \label{23}
\end{equation}
\begin{equation}
B_{s}=-[2{\tau}A_{s}+div{\vec{b}}A_{n}+({\kappa}+div{\vec{n}})A_{b}]\label{24}
\end{equation}
Algebraic manipulation of the above equations allow us to compute
the magnetic energy
\begin{equation}
E_{B}=\frac{1}{8{\pi}}\int{B^{2}dV} \label{25}
\end{equation}
which yields
\begin{equation}
E_{B}=\frac{1}{8{\pi}}\int{({B_{s}}^{2}+{B_{n}}^{2})dV} \label{26}
\end{equation}
or in terms in terms of the vector potential component $A_{n}$
yields
\begin{equation}
E_{B}=\frac{R^{2}{A_{n}}^{2}}{2}[\int{{\tau}^{2}ds}+\frac{1}{4}\int{div{\vec{b}}ds}]\label{27}
\end{equation}
which upon the assumption $div{\vec{b}}<<{\tau}^{2}$ reduces to
\begin{equation}
E_{B}=\frac{R^{2}{A_{n}}^{2}}{2}[\int{{\tau}^{2}ds}] \label{28}
\end{equation}
In case of helical filaments one knows that $\tau=c_{0}\kappa$ where
$c_{0}$ is constant and the expression (\ref{27}) reads
\begin{equation}
E_{B}=\frac{{c_{0}R}^{2}{A_{n}}^{2}}{2}[\int{{\kappa}^{2}ds}]
\label{29}
\end{equation}
Due to a theorem by Fukumoto \cite{13}
\begin{equation}
\int{{\kappa}^{2}ds}< 16{\pi}^{2}L \label{30}
\end{equation}
which clearly places a bound on the magnetic energy $E_{B}$. Here L
denotes the length of the filament. Algebraic manipulation of the
same equations yields a relation between the Frenet curvature and
torsion in terms of $div{\vec{n}}$ as
\begin{equation}
{\kappa}\pm\sqrt{2}{\tau}=-\frac{div{\vec{n}}}{2} \label{31}
\end{equation}
In the helical case we obtain the following equation
\begin{equation}
{\kappa}[1\pm\sqrt{2}]=-\frac{div{\vec{n}}}{2}  \label{32}
\end{equation}
Now let us compute the global helicity
\begin{equation}
H=\int{{\vec{A}}.\vec{B}dV} \label{33}
\end{equation}
which is equivalent to
\begin{equation}
H=\int{[{A_{s}}.{B_{s}}+{A_{n}}.{B_{n}}]dV}\label{34}
\end{equation}
substitution of the relations above yields
\begin{equation}
H=-\int{[2{\tau}A_{n}+div{\vec{b}}]A_{n}ds} \label{35}
\end{equation}
Due to a theorem which states
\begin{equation}
H=\frac{{\Phi}}{2{\pi}}[Tw+Wr]\label{36}
\end{equation}
 where ${\Phi}$ is the magnetic flux $\int{\vec{B}.d\vec{S}}$. Before computing the writhe number we need to compute the twist
of the magnetic filament. But this becomes quite simple thanks to a
formula recently derived by Berger and Prior \cite{5} which is given
by
\begin{equation}
\frac{d}{ds}Tw=\frac{J_{||}}{|\vec{B}|}=\frac{{\alpha}B_{s}}{B_{s}}=\alpha\label{37}
\end{equation}
where $J_{||}$ is the electric current along the twisted filament.
This result was obtained thanks to the assumption that
$B_{n}<<<B_{s}$ since $B_{n}$ is nothing but a simple perturbation
of $B_{s}$. Integration of the last expression yields
\begin{equation}
Tw=\int{{\tau}ds}\label{38}
\end{equation}
Substitution of the twist and magnetic helicity H into expression
(\ref{36}) yields finally the expression for the writhe number
\begin{equation}
Wr=-[(1+\frac{4{\pi}{A_{n}}^{2}}{{\Phi}^{2}})\int{{\tau}ds}+\frac{2{\pi}}{{\Phi}^{2}}(div\vec{b})\int{A_{n}ds}]\label{39}
\end{equation}
Note that the last term in this expression is proportional to the AB
phase $\int{A_{n}ds}$.
\section{Conclusions}
 In conclusion,a new expression for the writhe of magnetic vortex twisted filament has been obtained presenting a new term representing
 a Berry´s phase which was not present in the previous calculation of Moffatt and Ricca. A bound
 in the energy has been obtained thanks to a theorem by Fukumoto on
 the bound of the integral invariant of the square of the Frenet
 curvature value in the helical case. Future applications in plasma physics or in DNA \cite{12} be appear
 elsewhere. Expressions for the writhe number for the elastodynamics
 have been obtained by Klapper and Tabor \cite{13}. Other integral invariants in vortex fluid dynamics have been recently
 also computed by Maggioni and Ricca \cite{14} which can be generalised to MHD. Yet other interesting
 examples of the topological bounds to energy \cite{2} have been
 provided by Khesin \cite{15}.

 \section*{Acknowledgements}
 Thanks are due to CNPq and UERJ for financial supports.

\end{document}